\documentclass[%
 reprint,
superscriptaddress,
showpacs,
 amsmath,amssymb,
 aps,
floatfix,
]{revtex4-2}

\usepackage{mwe}
\usepackage{graphicx}
\usepackage{dcolumn}
\usepackage{bm}
\usepackage{hyperref}
\usepackage{color}
\usepackage[normalem]{ulem}
\usepackage{orcidlink}

\begin{document}

\title{Criticality at work: Scaling in the mouse cortex enhances performance}

\author{Gustavo G. Cambrainha \orcidlink{0009-0000-8153-305X} }
\author{Daniel M. Castro \orcidlink{0000-0002-2761-2133} }
\affiliation{Departamento de F{\'\i}sica, Centro de Ciências Exatas e da Natureza, Universidade Federal de Pernambuco,
Recife, PE, 50670-901, Brazil}

\author{Nivaldo~A.~P.~de~Vasconcelos~\orcidlink{0000-0002-0472-8205}}
\affiliation{Departamento de Engenharia Biom{\'e}dica, Centro de Tecnologia e Geociências, Universidade Federal de Pernambuco,
Recife, PE, 50670-901, Brazil}

\author{Pedro V. Carelli \orcidlink{0000-0002-5666-9606}}

\author{Mauro Copelli \orcidlink{0000-0001-7441-2858}}
\email{mauro.copelli@ufpe.br}
\affiliation{Departamento de F{\'\i}sica, Centro de Ciências Exatas e da Natureza, Universidade Federal de Pernambuco,
Recife, PE, 50670-901, Brazil}

\begin{abstract}
The critical brain hypothesis posits that neural systems operate near a phase transition, optimizing the processing of information. While scale invariance and non-Gaussian dynamics—hallmarks of criticality—have been observed in brain activity, a direct link between criticality and behavioral performance remains unexplored. 
Here, we use a phenomenological renormalization group approach to examine neuronal activity in the primary visual cortex of mice performing a visual recognition task. 
We show that nontrivial scaling in neuronal activity is associated with enhanced task performance, with pronounced scaling observed during successful task completion. 
When rewards were removed or non-natural stimuli presented, scaling signatures diminished. 
These results suggest that critical dynamics in the brain is crucial for optimizing behavioral outcomes, offering new insights into the functional role of criticality in cortical processing.

\end{abstract}
\maketitle

\section{Introduction}

The idea that living systems self-organize in ways that promote efficiency, resilience, and adaptation has captivated scientists across multiple fields~\cite{cavagna_scale-free_2010,mora_are_2011,hidalgo_information-based_2014,de_paoli_behavioral_2017,villegas_evidence_2024}.
In neuroscience, this notion is encapsulated by the critical brain hypothesis~\cite{beggs_neuronal_2003, beggs_criticality_2008, chialvo_emergent_2010, shew_functional_2013, PlenzNiebur14, TomenHerrmannErnst2019, munoz_colloquium_2018, obyrne_how_2022}, which posits that the brain operates near a phase transition, a tipping point between two different states. 
This regime would be able to maximize information transmission, information storage, learning and  dynamic range in response to incoming stimuli~\cite{bertschinger_real-time_2004,haldeman_critical_2005,kinouchi_optimal_2006,beggs_criticality_2008, shew_neuronal_2009,arcangelis_2010_learning,shew_information_2011,boedecker_information_2012,gautam_maximizing_2015,capano_2015_optimal,kessenich_2019_pattern,avramiea_long-range_2022,barzon_nicoletti_excitation}, leading to debates about the possible clinical implications of the hypothesis~\cite{zimmern_why_2020,breakspear_unifying_2006,dehghani_dynamic_2016,hohlefeld_longrange_2012,alamian_altered_2022}. 

One of the hallmark features of the critical point where a phase transition occurs is scale invariance: 
the system's activity has no characteristic length or time. 
Over two decades ago, this property was initially found in the self-similar temporal structure of human electroencephalogram (EEG) data~\cite{linkenkaer-hansen_long-range_2001} and in the scale-free nature of neuronal avalanches---bursts of electrical activity---in cortical slices~\cite{beggs_neuronal_2003}. 
Since then, a growing body of evidence supporting the existence of the putative phase transition has been gathered in different experimental setups~\cite{ribeiro_spike_2010,lombardi_balance_2012,tagliazucchi_criticality_2012,palva_neuronal_2013,shriki_neuronal_2013,fontenele_criticality_2019,dahmen_second_2019,ma_cortical_2019,jones_shew_scalefree}. 

Recently, a novel phenomenological renormalization group (PRG) method was introduced~\cite{bradde_pca_2017,meshulam_coarse_2019}, offering a fresh approach to studying scale-invariant behavior in complex systems. 
Inspired by the renormalization group (RG)—the gold standard for understanding phase transitions in theoretical models~\cite{kadanoff_scaling_1966,wilson_renormalization_1971,wilson_renormalization_1971-1,fisher_renormalization_1998}—the PRG similarly reveals nontrivial scale invariance at non-Gaussian fixed points. 
It does so by tracking how observables evolve across scales as the system undergoes recursive coarse-graining. However, unlike the RG, the PRG is model-free and directly applicable to high-dimensional data from complex systems, making it particularly useful in neuroscience, where experimental advancements now enable the simultaneous recording of thousands of neurons~\cite{jun_fully_2017}.
The effectiveness of the PRG method has been validated through various studies, ranging from neural activity in \textit{C. Elegans} to the whole human brain~\cite{meshulam_coarse_2019, morales_quasiuniversal_2023, ponce-alvarez_critical_2023, castro_and_2024, munn_multiscale_org}.

Despite the increasing methodological sophistication in the methods to assess scale invariance in neuronal data, a glaring issue remains. 
A link between criticality and task performance---the alleged reason why critical dynamics should be found in the brain in the first place---has never been found. Here we bridge this gap, showing that nontrivial scaling in the primary visual cortex of mice is associated with enhanced performance in a visual task.


\begin{figure*}[ht!]
\centering
\includegraphics[width=0.9\textwidth]{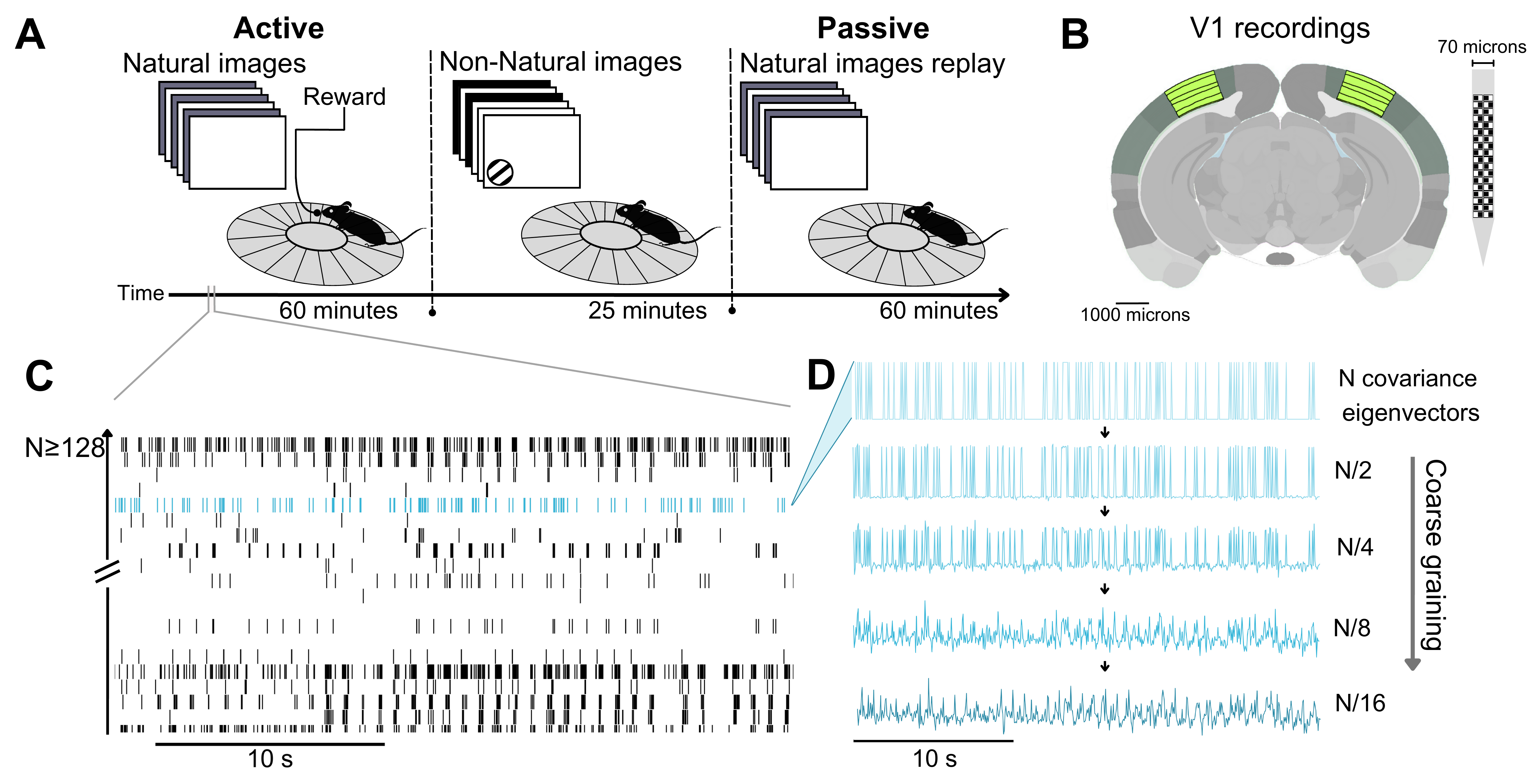}
\caption{The phenomenological renormalization group applied to mouse brain data.
(A) 
In the first part (Active) of the Allen Brain Observatory Visual Behavior Neuropixels experiment~\cite{whitepaper}, natural images are shown in 250-ms flashes and the mouse needs to recognize a change of images in order to receive a reward. 
In sequence, the screen shows non-natural images such as Gabors, along with black and white flashes and a gray background. 
In the third part (Passive), the natural images are repeated in the same order as in the first part, but the mouse can no longer  lick to be rewarded.
(B) 
Frontal slice of a mouse brain with the primary visual cortex (V1) highlighted in bright green (left) 
(credits to Allen Institute for Brain Science) 
and scheme of a Neuropixels 1.0 probe (right) used in the data collection. 
(C) 
Raster plot of spiking activity for a 30-s window.
(D)
Progressive coarse-graining of a unit time series, corresponding to projections onto a decreasing number of eigenvectors of the covariance matrix ($N, N/2, N/4,\ldots$). 
The variance of the activity is normalized to 1.
}
\label{fig1}
\end{figure*}

\begin{figure*}[ht!]
\centering
\includegraphics[width=0.9\textwidth]{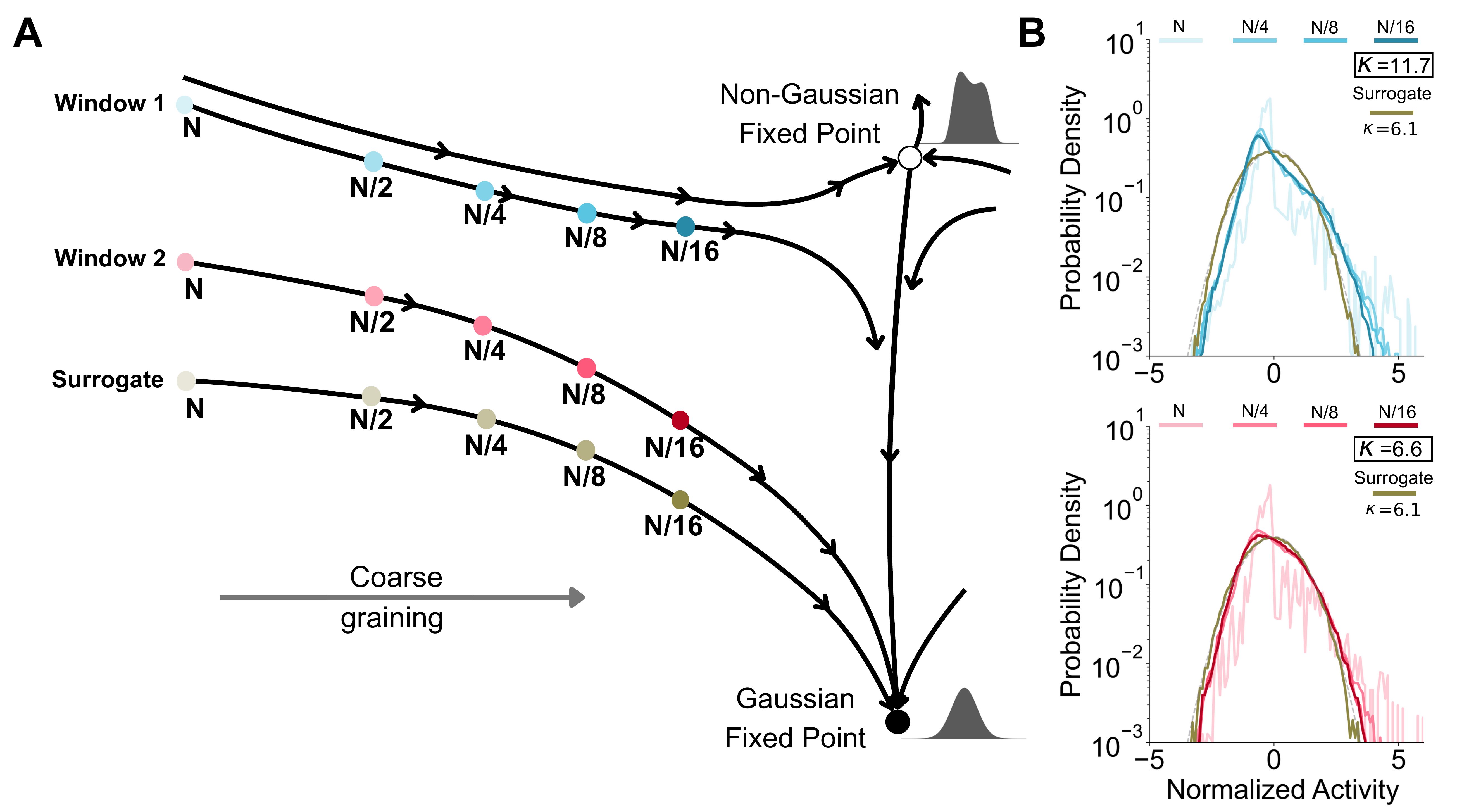}
\caption{The phenomenological renormalization group applied to mouse brain data.
(A)
Schematic representation of the renormalization flow created by the coarse-graining procedure. 
Time windows that are further from criticality, such as those from spike-shuffled surrogate data, converge more rapidly to the Gaussian fixed point. 
(B)
In the final step of the coarse-graining procedure, windows closer to criticality (blue) retain their non-Gaussian shape, with higher kurtosis, while others (red) come closer to a Gaussian distribution (dashed line), with lower kurtosis (see Fig.~S6 for the evolution of the distribution shape with kurtosis).
}
\label{fig12}
\end{figure*}


\section{Results}\label{results}

\subsection{Performance increases with kurtosis}
Since our goal is to understand the relation between brain criticality and behavior, we analyzed neuron spiking activity from the primary visual cortex (V1) of mice while they performed a visual task~\cite{Siegle_Jia_Durand_Gale_Bennett_Graddis_Heller_Ramirez_Choi_Luviano_et} [Figs.~\ref{fig1}A-\ref{fig1}C]. 
This activity was subdivided into 30-s  windows (see the Appendix). 
For each window, we calculated the covariance matrix $C_{ij}$ comprising all pairs of recorded neurons. 
By diagonalizing this matrix, we obtained a set of eigenvectors from which projectors were created.
When projecting the activity of each neuron over all eigenvectors, one recovers the original time series. 
By progressively selecting fewer components corresponding to the largest eigenvalues for projection, we generated an increasingly coarse-grained representation of the neural activity (a procedure akin to discarding faster modes in momentum space in standard RG)~\cite{bradde_pca_2017} [Fig.~\ref{fig1}D]. 
We tracked the normalized distribution of this activity as we chose fewer covariance eigenvectors. 
This procedure creates a flow of activity distributions [Figs.~\ref{fig12}A~and~\ref{fig12}B] that pushes noncritical behavior towards a trivial (Gaussian) fixed point.
For each window, we used the kurtosis $\kappa$ of the distribution in the final coarse-graining step as a metric to quantify the deviation from the Gaussian distribution.

Certain time windows exhibit clear signatures of nontrivial dynamics, as expected at a critical point, with kurtosis values significantly higher than those in the spike-shuffled (surrogate) data. 
In contrast, other windows show results similar to surrogate data [Fig.~\ref{fig12}B].
Across 62 sessions (from 44 mice), we observed a broad range of kurtosis values, overlapping with the distribution from the surrogate data [Fig.~\ref{fig2}A]. 
As we depart from the Gaussian distribution, the distinction between windows with trivial and nontrivial statistics is arbitrary.
We exploit this arbitrariness by establishing a kurtosis threshold $\kappa_{\textrm{th}}$: a window with $\kappa > \kappa_{\textrm{th}}$ is deemed nontrivial, with scaling properties; otherwise, it reflects Gaussian activity without scaling [Fig.~\ref{fig2}A]. 
We allow $\kappa_{\textrm{th}}$ to assume any value greater than zero. 

\begin{figure*}
    \centering
    \includegraphics[width=\linewidth]{figure2k.pdf}
    \caption{Improved task performance is linked to higher kurtoses. 
(A) Distribution of kurtosis values at the final coarse-graining step (projections onto $N/16$ covariance eigenvectors) across time windows from all subjects (aggregated data). 
Unlike the surrogate data, the kurtosis of real data span a wider range of values. 
Windows with kurtosis values above (below) a tunable threshold ($\kappa_{\textrm{th}}$) are labeled as scaling (no-scaling) windows.
(B)
The performance of real data is compared to data where hits and misses are shuffled (trial shuffled). 
(C)
The hit rate calculated across scaling windows is higher than both no-scaling windows and trial-shuffled data for all kurtoses exceeding the maximum value obtained for surrogate data $\kappa_{\textrm{max}}^{\textrm{surr}}$ (shaded area). 
The dashed line shows the kurtosis distribution. 
(D) The relative performance gain is calculated from the difference between scaling and no-scaling curves, showing up to a $24\%\pm3\%$ increase in performance.
}
\label{fig2}
\end{figure*}    

The experiment has three phases.
During the first phase, the mice perform a task of image change recognition every few seconds, and they are rewarded for correct responses [hits, Fig.~\ref{fig2}B]~\cite{whitepaper}.
For any value of $\kappa_{\textrm{th}}$ above surrogate values, the average hit rate was consistently higher when calculated over scaling windows than over no-scaling windows [Fig.~\ref{fig2}C].
We have verified that higher kurtosis was associated to better performance [Fig.~\ref{fig2}C], with relative performance gain reaching up to $24\% \pm 3\%$  [Fig. \ref{fig2}D].
As a control, we confirmed that the results were better than those obtained when hits and misses were shuffled within each session [trial-shuffled data, Fig.~\ref{fig2}B].
Time windows of 10 and 60~s were also analyzed and showed similar results, associating higher kurtosis with better performance (see the Supplemental Material).
\begin{figure*}
    \centering
\includegraphics[width=0.9\linewidth]{figure3f.pdf}
\caption{Comparison of the three phases of the experiment.
(A) Fraction of windows above a given kurtosis threshold ($\kappa_{\textrm{th}}$) in the aggregated data (complementary cumulative distribution function). 
The active phase remains consistently above the other phases for any choice of $\kappa_{\textrm{th}}$, indicating a less Gaussian-like behavior. 
The shaded area corresponds to kurtoses of surrogate data.
(B-E) Comparison of scaling exponents ($\alpha$, $\beta$, $\mu$ and $z$) during the three phases. 
Bars (lines) represent the mean values (standard deviations) across sessions. 
Lighter shades depict results from surrogate data. 
Dashed lines correspond to the trivial values of exponents $\alpha$, $\beta$ and $z$.
Triple asterisks (*) indicate a $p$-value $<0.001$ for the statistical comparison between two distributions utilizing the Mann-Whitney U test.}\label{fig3}
\end{figure*}

\subsection{Signatures of criticality change with context}
In the passive phase, the same images are shown, but there is no task to be performed or reward to be gained (the lick port is removed). 
During this phase, the distribution of kurtoses across windows got closer to surrogate values than in the active phase. 
For any $\kappa_{\textrm{th}}$, scaling windows were more abundant in the active phase than in the passive phase [Fig.~\ref{fig3}A].
Before the passive phase, mice were presented with non-natural images, such as Gabor patches, black and white screen flashes, or a gray background~\cite{whitepaper}. 
In this control condition, the shift towards lower levels in the kurtosis distribution was even more pronounced, and scaling windows became less likely to occur 
[though most windows still remained with higher kurtosis than those of surrogate data, Fig.~\ref{fig3}A].
These findings indicate that the process of performing a task drives an adaptation in the dynamics of the visual cortex, directing the system toward a non-Gaussian fixed point.
Two additional analyses are available in the Supplemental Material. 
The first investigates to which extent the population firing rate relates to kurtosis and performance across experimental conditions.
The second focuses on the possible effects of fatigue: in the second half of the active phase, we observe lower kurtoses than in the first half, but the relation between performance and kurtosis is preserved.

We also calculated phenomenological renormalization scaling exponents using a real-space approach to the coarse-graining method~\cite{meshulam_coarse_2019,nicoletti_scaling_2020}. 
This process begins by calculating pairwise correlations from neuronal activity and progressively summing the time series of the most correlated pairs of neurons until there are none left unpaired. By iterating this process $k$ times, we obtain coarse-grained variables (or “clusters”) that are sums of $2^k$ spiking time series.

Near a non-Gaussian fixed point, several quantities will scale with cluster size following nontrivial power laws, namely the mean variance (exponent $\alpha$), minus the log of the probability of silence within a cluster ($\beta$), and the mean autocorrelation characteristic time ($z$).
Additionally, the covariance eigenvalues scale with their rank with exponent $\mu$~\cite{meshulam_coarse_2019, nicoletti_scaling_2020, morales_quasiuniversal_2023, ponce-alvarez_critical_2023, castro_and_2024} (see Appendix \ref{renorm}). 
For the active phase, we obtained $\alpha_1=1.41\pm0.05$, $\beta_1=0.77\pm0.04$, $\mu_1=0.73\pm0.08$ and  $z_1=0.3\pm0.1$ [Figs.~\ref{fig3}B-\ref{fig3}E]. 
These scaling exponents show minimal change when calculated for kurtoses above \(\kappa_{\textrm{th}}\), with only a weak dependence on this threshold (see the Supplemental Material).
During the passive phase we found $\alpha_2={1.37\pm0.04}$, $\beta_2=0.78\pm0.04$, $\mu_2=0.61\pm0.05$, $z_2=0.36\pm0.06$, with statistically significant differences with respect to the active phase for $\alpha$, $z$ and $\mu$  [Figs.~\ref{fig3}B-\ref{fig3}E]. 
These values align with those reported for awake mice in a different dataset~\cite{morales_quasiuniversal_2023}.
Furthermore, for non-natural stimuli the exponents were $\alpha_3=1.28\pm0.03$, $\beta_3=0.82\pm0.03$, $\mu_3=0.38\pm0.04$ and $z_3=0.4\pm0.1$, which are all different from the phases with natural images [Figs.~\ref{fig3}B-\ref{fig3}E].

\section{Discussion}

We observed that non-Gaussian scaling dynamics in the mouse primary visual cortex is closely linked with improved performance in a visual task, supporting the idea that critical dynamics underlies functional advantages in neural processing.
These findings provide strong evidence for the critical brain hypothesis, which suggests that neural systems operate near a critical point to optimize information processing and may be essential for understanding brain function~\cite{mora_are_2011,obyrne_how_2022}.

The kurtosis of the coarse-grained activity distribution offers a powerful means of assessing critical dynamics, with its flexibility---particularly the tunable threshold (\(\kappa_{\textrm{th}}\))---enabling  control over which windows are classified as ``truly'' scaling. 
In essence, kurtosis acts as a \textit{de facto} measure of the ``distance to triviality.''
Our findings show that as we consider only windows of progressively higher kurtosis (closer to the critical state), the corresponding performance calculated across those windows also rises.
By increasing \(\kappa_{\textrm{th}}\) to focus on only the highest kurtosis values, performance gains in scaling windows intensify, reaching up to $24\%\pm 3\%$ over non-scaling periods [Figs.~\ref{fig2}C and~\ref{fig2}D]. 
 
Moreover, we found that the signatures of criticality in V1 are modulated by the level of task involvement.
For the exact same set of images presented in the same sequence, kurtosis levels observed in the active phase drop considerably in the passive phase, simply by preventing the mouse from actively engaging with them.
When presented with non-natural images, kurtoses drop even further toward trivial values. 
The mouse brain's ability to self-regulate and display scaling dynamics predominantly while performing a task reinforces the idea that scaling plays a role in enhancing sensory processing. 

The four scaling exponents also reflect the differences among the phases of the experiment. 
When animals are shown natural images and are actively performing the task, $\alpha$ and $\beta$ (``static exponents"~\cite{munn_multiscale_org}) deviate significantly from trivial values $\alpha=1$ and $\beta=1$ [Figs. \ref{fig3}B and~\ref{fig3}C]. 
In contrast, during a passive replay of natural images, the static exponents shift closer to trivial values, approaching this baseline even more closely when non-natural images are used as stimuli without any task involved. 
The ``dynamic exponent" $z$ exhibited an inverse trend, diverging progressively from the trivial value $z = 0$ across phases: increasing from active to passive and peaking in the non-natural image phase. 
This pattern is reminiscent of findings from a recently developed temporal renormalization group technique, which showed a deviation from critical behavior during action compared to deep sleep~\cite{sooter_cortex_2024}.
Furthermore, the measured values of the exponent $\mu$ reinforce the idea that these differences are not purely input-driven~\cite{stringer_2019_nature}. 
Instead, the observed values align more closely with background activity (orthogonal to the input), as reported in a recent study~\cite{morales_quasiuniversal_2023}.
Together, these findings provide new insights into the relationship between behavior and criticality in the primary visual cortex, suggesting that task engagement and stimulus type modulate cortical scaling properties.

The spread in both exponent and kurtosis values observed in Fig.~\ref{fig3} is consistent with a recent computational study using this technique~\cite{nasciment2025arXiv}.
That study reports that PRG produces a continuum of values for these quantities near a phase transition, with a distinct peak appearing only at the critical point.
Taken together with our results, this suggests that the brain may operate not exactly at criticality, but rather in a slightly off-critical regime (such as a subcritical state), 
a view consistent with previous discussions in the literature~\cite{Priesemann2014subcriticalbrain}.

Since pioneering experiments first reported neuronal avalanches in neural systems, numerous theoretical studies have proposed that the brain’s operation near criticality confers functional advantages, enhancing processing efficiency and adaptability. 
Although earlier  studies examined these effects in cortical slices~\cite{beggs_neuronal_2003}, anesthetized rats~\cite{gautam_maximizing_2015}, or explored the functional role of criticality in health-related contexts~\cite{hengen_shew_is_criticality}, no prior work has directly associated neuronal criticality with functional performance in a living, awake, and behaving animal, nor shown how these signatures of criticality change with task engagement.
Though we are not able to provide a biophysical mechanism by which brain criticality enhances performance, our research establishes a phenomenological link between the critical brain hypothesis and its main flagship: that criticality underlies the brain's efficient dynamics, facilitating optimized behavior and cognition. 
Furthermore, our finding that scaling properties depend on the animal's context opens new avenues for investigating how criticality might contribute to diverse brain functions and exploring the underlying mechanisms that sustain this state across varying cognitive demands and environmental challenges.

\begin{acknowledgments}
The authors acknowledge support from Brazilian agencies CNPq (Grants No.~140660/2022-4, No.~308703/2022-7, No.~314094/2023-7, No.~444500/2024-3 and No.~408389/2024-9), FACEPE (Grant No.~APQ-1187-1.05/22), FINEP (Grant No.~NeuroAssist~01.24.0124.00) and FADE/UFPE (Grant
No.~64/2024).

M.C. and D.M.C. designed the study.
D.M.C. provided initial code for the analysis.
G.G.C. refined the code, ran the analysis and made the plots.
All authors discussed the results and wrote the paper. 
\end{acknowledgments}

\section*{DATA AVAILABILITY}
The data that support the findings of this article are openly available~\cite{allendandi}.
\appendix
\section*{Appendix: Methods and Materials}\label{materiaslandmethods}
\subsection{Data}\label{secdata}
The data used in this study come from the Allen Visual Behavior Neuropixels dataset~\cite{allendandi}, which includes neuronal activity recorded from the V1 region of mice using Neuropixels 1.0 probes. Each recording spans approximately 2 h and 25 min.
Prior to the experiment, mice were trained to lick a device upon detecting an image change in order to receive a reward. During the first hour of the experiment, they were shown a series of images under these trained conditions.
Following this phase, mice were shown Gabor patches for 15 min, then exposed to black and white flashes, and finally shown a gray screen for 5 min. 
In the third phase, the natural images are shown a second time in the same order as before, but this time the lick apparatus is retracted.
The raw electrophysiological data were preprocessed and sorted into spike trains using the Kilosort2 algorithm~\cite{whitepaper}. 
We selected all 62 sessions from the dataset that had more than 128 recorded units, which included both single-unit activity (SUA) and multiunit activity (MUA). 
The data were then divided into these three experimental phases, binned into 50-ms intervals, and segmented into 30-s windows for analysis.

We excluded from the analysis temporal windows based on the coefficient of variance (CV)
\begin{equation}
\textrm{CV}=\frac{\sigma}{\mu},
\end{equation}
where $\mu$ and $\sigma$ are the mean and the standard deviation of the activity across the window.
Windows where $\textrm{CV}>5$ were excluded, as such high values are not typically observed in awake mice and likely result from the chosen window size, indicating artefacts.

\subsection{Surrogate data}
Throughout our analysis, we compared our results with surrogate data. 
These surrogate data were generated by shuffling the interspike intervals of each unit within each time window, effectively breaking correlations between units and producing trivial kurtosis values and exponents for comparison.

\subsection{Momentum-space renormalization and kurtosis}
The coarse graining procedure initially proposed by Bradde and Bialek~\cite{bradde_pca_2017} was applied to each of the 30-s time windows. Here we give a brief account of the method.

We start from the covariance matrix
\begin{equation}\label{cov}
    C_{ij}=\langle \phi_i\phi_j\rangle-\langle\phi_i\rangle\langle\phi_j\rangle,
\end{equation}
where $\phi_i$ is the time series activity of unit $i$. 
Taking its ranked eigenvalues ${\lambda_1 > \lambda_2 > \cdots > \lambda_N}$ 
and the components $u_{\mu i}$ of their respective eigenvectors,  such that
\begin{equation}
    \sum^{N}_j C_{ij}u_{\mu j}=\lambda_\mu u_{\mu i}\; ,
\end{equation}
we write the projector~\cite{bradde_pca_2017}
\begin{equation}\label{equ3}
    \hat{P}_{ij}(N_\mathrm{cutoff})=\sum^{N_\mathrm{cutoff}}_{\mu=1}u_{\mu i}u_{\mu j}.
\end{equation}
Using Eq.~\eqref{equ3} we obtain a set of variables $\psi_i$ for a chosen value of $N_{\mathrm{cutoff}}$
\begin{equation}
    \psi_{i}(N_{\mathrm{cutoff}})=\mathcal{Z}_i(N_{\mathrm{cutoff}})\sum^{N}_{j=1}\hat{P}_{ij}(N_{\mathrm{cutoff}})\left(\phi_i-\langle \phi_i\rangle\right)\; ,
\end{equation}
where $\mathcal{Z}_i(N_{\mathrm{cutoff}})$ is such that $\text{var}(\psi_{i})=1$, 
as shown in Fig.~\ref{fig1}D for $N_{\mathrm{cutoff}} = N$, $N/2$, $N/4$, $N/8$, $N/16$~\cite{nicoletti_scaling_2020,castro_and_2024}.

From the distribution of $\phi_i$ at the final coarse graining step ($N_{\mathrm{cutoff}} = N/16$), we  calculate the kurtosis $\kappa=\langle\psi^4\rangle/\langle\psi^2\rangle^2$ [Fig.~\ref{fig1}F]. 
Non-critical dynamics converges towards a Gaussian distribution ($\kappa=3$) as we reduce the number of eigenvectors involved in the construction of the projector [Eq.~\ref{equ3}], while critical dynamics does not.
Note that coarse-grained distributions do not converge perfectly to a Gaussian even for surrogate data (with the lowest kurtoses around $\kappa\simeq 6$), due to the finite number of bins within a window.

\subsection{Real-space renormalization and exponents}\label{renorm}

To obtain the PRG scaling exponents presented in Figs.~\ref{fig3}B-\ref{fig3}E we use a different phenomenological renormalization process, which we call real-space renormalization~\cite{meshulam_coarse_2019}. 
In this analysis we cluster together the activities of the most correlated pair of neurons repeatedly until we exhaust all neurons: 
\begin{equation}
\sigma_i'^{(k+1)}=\sigma_i^{(k)}+\sigma_j^{(k)},
\end{equation}
where $j$ is the most correlated neuron to $i$. 
This procedure is repeated $k$ times, after which we have $N/2^k$ coarse-grained units (or ``clusters''), each comprising the sum of $K=2^k$ original spiking units. 
We follow some quantities that are expected to scale with nontrivial power laws only close to the critical regime.
\paragraph{\normalfont \textit{Mean variance}}
$M_2$, the variance of the activity of the clusters of size $K$, averaged over all clusters, grows with the cluster size as
\begin{equation}
    M_2\propto K^\alpha.
\end{equation}
Uncorrelated neurons lead to the trivial value $\alpha=1$ via the central limit theorem. 
Perfectly correlated neurons lead to $\alpha=2$. 

\paragraph{\normalfont \textit{Silence probability}}

An effective free-energy can be written using the probability $P_{\mathrm{silence}}$ of finding a silent bin within each cluster:  
\begin{equation}
F=-\log(\mathcal{P_{\mathrm{silence}}})\propto K^\beta.
\end{equation}
For the trivial case of uncorrelated Poisson spiking units, one obtains $\beta=1$. 

\paragraph{\normalfont \textit{Covariance spectrum}}

The spectrum of eigenvalues of the covariance matrix [Eq.~\ref{cov}] when ordered in descending order should follow

\begin{equation}
\lambda =A\left( \frac{K}{\mathrm{rank}} \right) ^\mu.
\end{equation}
For trivial data, this scaling is broken and $\mu$ is undefined. Time windows with a fit quality of $r^2<0.85$ for this power-law have their corresponding exponent discarded, as it cannot be reliably considered indicative of scaling.

\paragraph{\normalfont \textit{Temporal correlation}}

Let $C(t)$ be the average autocorrelation function $C(t)$ across clusters of size $K$, from which we take the characteristic autocorrelation time $\tau_c$, defined by $C(\tau_c)=1/e$.
This characteristic time scales with the cluster size as
\begin{equation}
    \tau_c\propto K^z\; .
\end{equation}
For the trivial case of uncorrelated spike trains, one has $z=0$.

\subsection{Hit rate}\label{sechitrate}
During the first phase of the  Allen Brain Observatory Visual Behavior Neuropixels experiment, mice performed a visual recognition task. 
To assess the impact of criticality on performance, we subdivided the data into 30-s intervals. 
After applying the PRG procedure streamlined in the previous section, intervals with a kurtosis above a chosen $\kappa_{\textrm{th}}$ were labeled as scaling windows.
Within each 30-s window only three to five trials are performed, leading to poor statistical estimation of performance on a window-to-window basis.
For this reason we chose to calculate the hit rate for these scaling windows by summing all correctly performed tasks (hits) within them and dividing by the total number of tasks in these intervals,
\begin{widetext}
\begin{align}
    \mathrm{Hit~rate~(Scaling)}&=\frac{ \mathrm{Total~number~of~sucessful~tasks~(windows~with~\kappa> }\kappa_{\textrm{th}})}{\mathrm{Total~number~of~tasks~(windows~with~\kappa>}\kappa_{\textrm{th}})}, \\
    \mathrm{Hit~rate~(No~scaling)}&=\frac{ \mathrm{Total~number~of~sucessful~tasks~(windows~with~\kappa<}\kappa{_\textrm{th}})}{\mathrm{Total~number~of~tasks~(windows~with~\kappa<}\kappa_{\textrm{th}})}.
\end{align}
\end{widetext}
The standard error of the mean was estimated through a bootstrapping statistic method by choosing $\mathcal{N}$ samples with replacement from the subset of time windows above (below) the kurtosis threshold, averaging and taking the standard deviation over $n=10^5$ repetitions.

To quantify the enhancement in performance due to scaling over no-scaling windows, we define the relative performance gain as 
\begin{widetext}
\begin{equation}
    \mathrm{Relative~performance}=\frac{\mathrm{Hit~rate~(Scaling)}-\mathrm{Hit~rate~(No~scaling)}}{\mathrm{Hit~rate~(No~scaling)}}.
\end{equation}
    
\end{widetext}

\subsection{Trial-shuffled data}

To control for artifacts, we generated a trial-shuffled version of the data. 
For each trial, we randomly selected another trial within the same session and swapped their outcomes [hit or miss, Fig.~\ref{fig2}B]. 
This process preserves the total number of hits and misses within each session, as well as the total number of trials within each window. 

For the trial-shuffled data, we calculated the hit rate for the scaling windows as described in the previous section, repeating this procedure 500 times. 
The mean and standard deviation across these repetitions are presented in Fig.~\ref{fig2}C.

\bibliography{copelli}

\end{document}


\title{Supplemental Material \\ Criticality at work: Scaling in the mouse cortex enhances performance}

\author{Gustavo G. Cambrainha \orcidlink{0009-0000-8153-305X} }
\author{Daniel M. Castro \orcidlink{0000-0002-2761-2133} }
\affiliation{Departamento de F{\'\i}sica, Centro de Ciências Exatas e da Natureza, Universidade Federal de Pernambuco,
Recife, PE, 50670-901, Brazil}

\author{Nivaldo~A.~P.~de~Vasconcelos~\orcidlink{0000-0002-0472-8205}}
\affiliation{Departamento de Engenharia Biom{\'e}dica, Centro de Tecnologia e Geociências, Universidade Federal de Pernambuco,
Recife, PE, 50670-901, Brazil}


\author{Pedro V. Carelli \orcidlink{0000-0002-5666-9606}}

\author{Mauro Copelli \orcidlink{0000-0001-7441-2858}}
\email{mauro.copelli@ufpe.br}
\affiliation{Departamento de F{\'\i}sica, Centro de Ciências Exatas e da Natureza, Universidade Federal de Pernambuco,
Recife, PE, 50670-901, Brazil}

\maketitle



\subsection*{Supplemental Text}



\section{Effects of different time window lengths}

Our choice of time window length to perform the analysis was made to balance two competing considerations: on the one hand, to have a sufficiently long time for the surrogate procedure to effectively shuffle the data within each window; on the other hand, to avoid an overly broad time frame that could blend different cortical dynamics. 

To examine the sensitivity of our results to this choice of window length, we repeated the analyses using 10-s and 60-s windows. 
This allowed us to evaluate the impact of shorter and longer time windows on our performance results.

\begin{figure*}[!h]
\centering
\includegraphics[width=\textwidth]{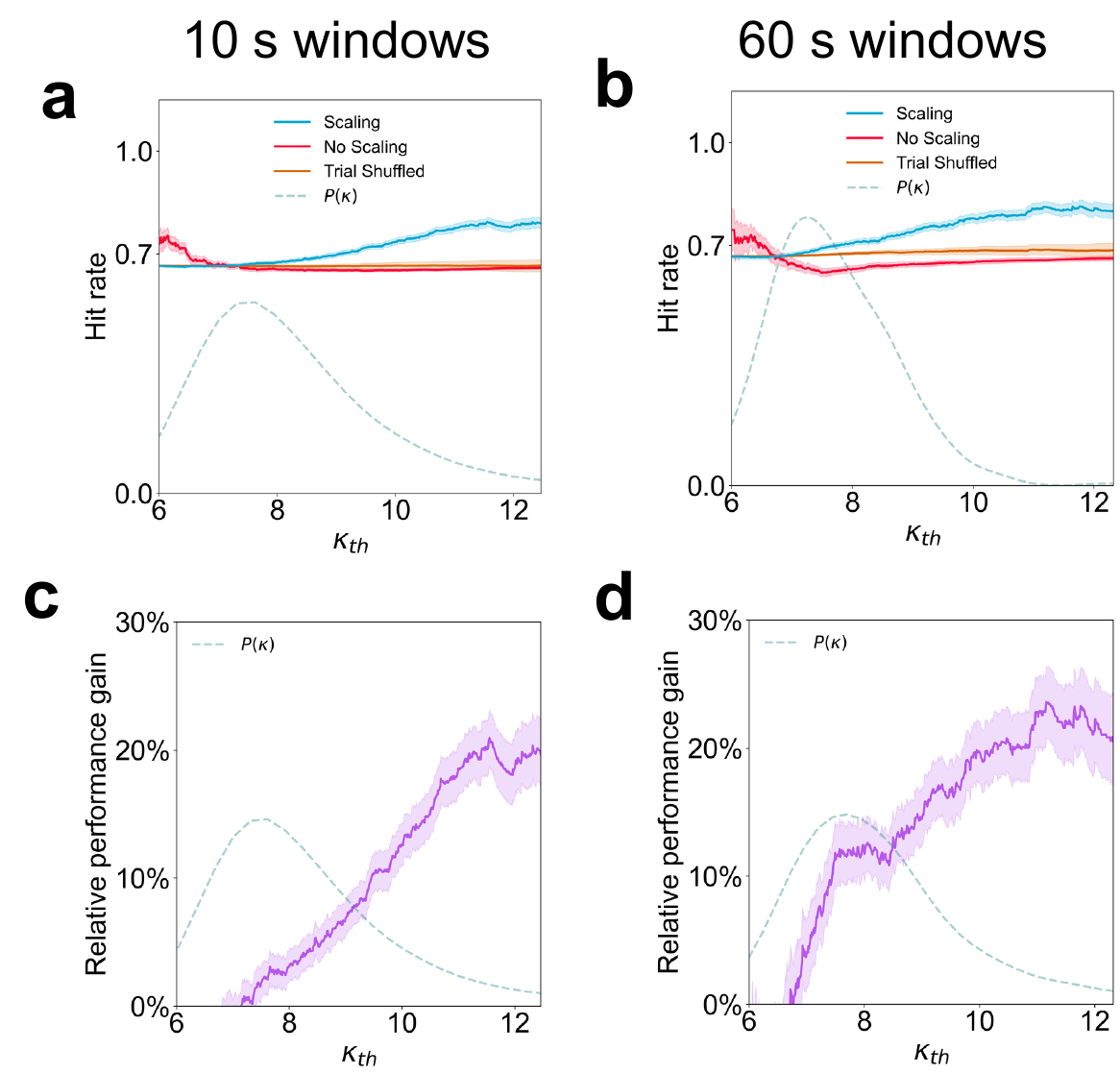}
\caption{\textbf{Relation of task performance and kurtosis for different windows sizes.} 
\textbf{(A-B),} hit rates for the 10-s and 60-s windows exhibit a similar trend to those of 30-s windows, with increasing hit rates as the threshold ($\kappa_{th}$) is increased. 
\textbf{(c-d),} Relative performance follows a similar pattern, deviating from zero around $\kappa\approx7$.}
\label{figSMwin}
\end{figure*}

We observe that the change in window size choice does not qualitatively alter our results (Fig.~\ref{figSMwin}). We consistently observe that windows with higher kurtosis are associated with improved performance. 
In Fig.~\ref{figSMwin}, the kurtosis range is omitted because the limited window size results in insufficient ISI shuffling, making it unreliable.

\section{Presence of noise units}

A natural question arises regarding whether our choice to use a greedy spike sorting approach (potentially including noise units) could influence the results. 
To address this, we repeated the analyses using only the 43 sessions that retained more than 128 units after applying more strict sorting criteria, while maintaining the 30-s time window.

\begin{figure*}[h!]
\centering
\includegraphics[width=\textwidth]{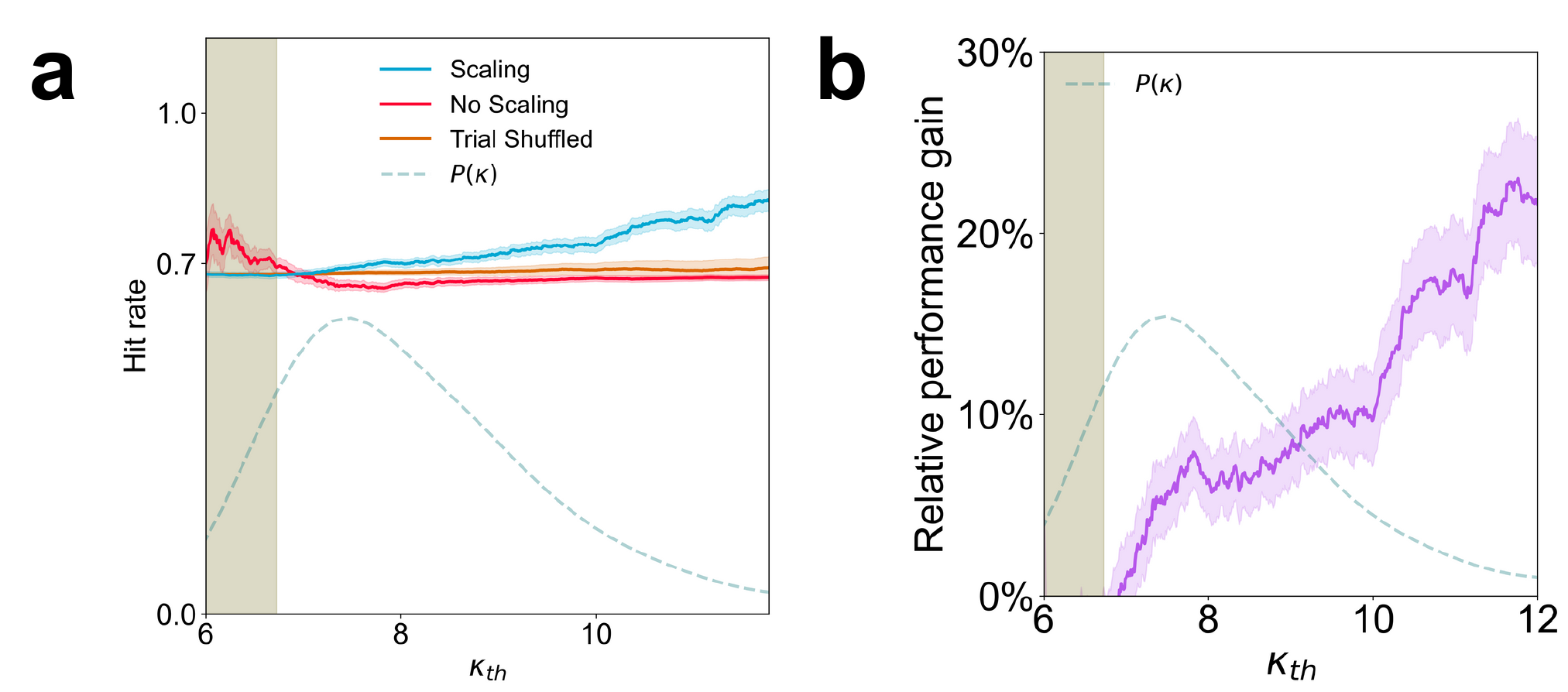}
\caption{\textbf{Relation of task performance and kurtosis for only ``no noise'' units.} \textbf{(A-B)} We repeat the plots presented in the main text, the results of increased performance are kept.}\label{fignonoise}
\end{figure*}

The results remained qualitatively consistent with those observed in the initial analyses (Fig.~\ref{fignonoise}). This outcome is expected, as a procedure in the spirit of the renormalization group, which captures large-scale properties, should be resilient to the inclusion of noise units.

\section{Exponents vs. threshold}

The exponents presented in Fig.~3 are calculated across all windows of a given session. However, it is reasonable to question this approach, as only a subset of these windows may truly exhibit scaling behavior. 
To address this, we analyze how the exponents $\alpha$, $\beta$, $\mu$ and $z$  change when calculated using only the windows with $\kappa>\kappa_{th}$. 

\begin{figure*}[h!]
\centering
\includegraphics[width=\textwidth]{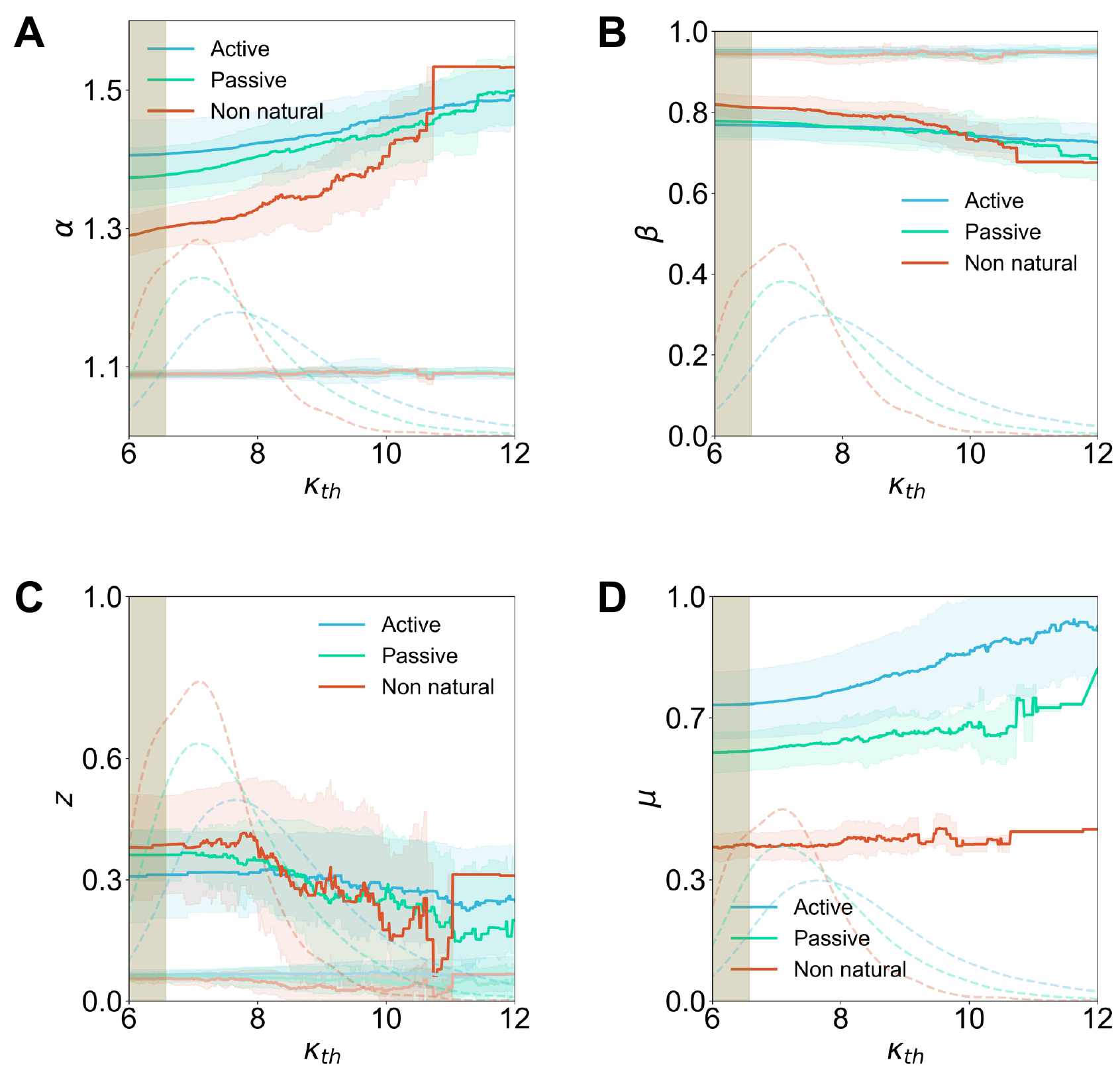}
\caption{\textbf{Exponents variation with $\kappa_{th}$.} \textbf{(A-D)} Exponent values show only minimal variation with different $\kappa_{th}$. 
The light solid lines represent the surrogate data, 
while dashed lines show the kurtosis distribution $P(\kappa)$ of each experimental phase.}\label{figexpkth}
\end{figure*}

We find that the exponent values vary only slightly and remain within their respective error bars (Fig.~\ref{figexpkth}). The larger fluctuations for larger kurtosis values are due to the small number of windows (note that $P(\kappa)$ has decreased considerably in those regions). 

\section{Activity kurtosis vs. coarse-grained activity kurtosis}

\begin{figure*}[h!]
\centering
\includegraphics[width=\textwidth]{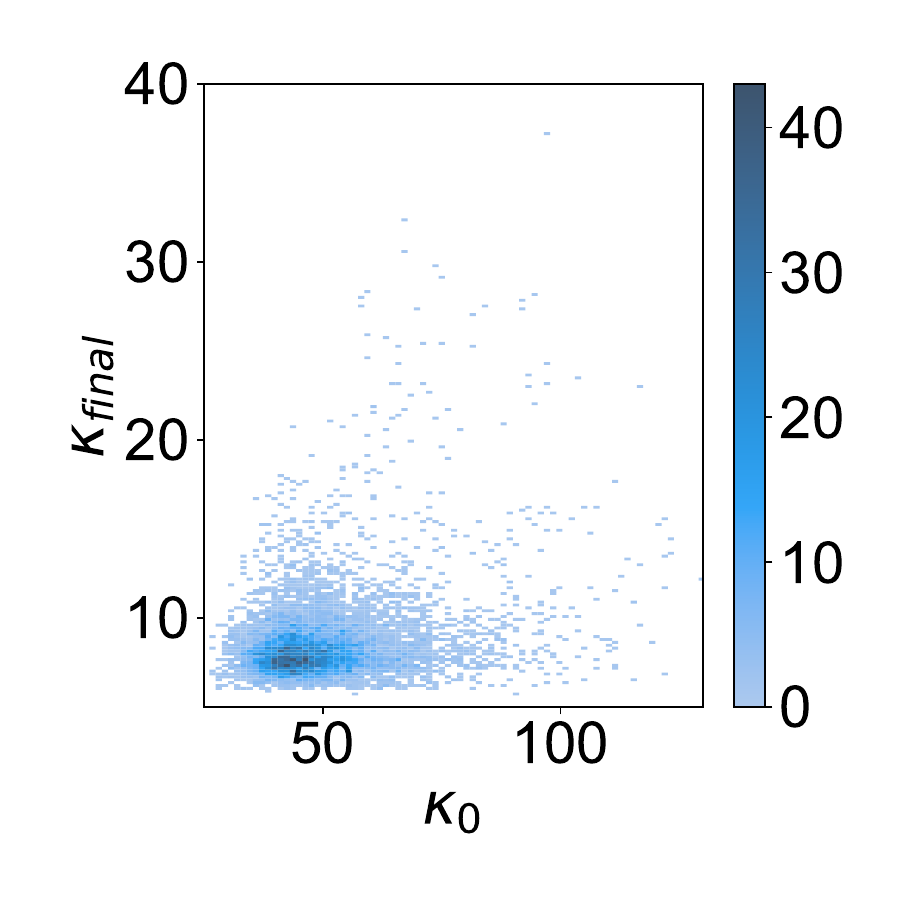}
\caption{\textbf{Comparison of initial and final kurtoses.} For each window we plot the value of the kurtosis of the activity distribution after the coarse-graining ($\kappa_{\mathrm{final}}$) against their initial kurtosis from the activity distribution ($\kappa_0$). 
There is no clear tendency in the relation between those two variables.}\label{figkappakappa}
\end{figure*}

One could wonder whether coarse-graining the neural activity is necessary to compute the kurtosis, or if examining the kurtosis of the original, non-coarse-grained activity distribution would suffice.
Our analysis shows that the initial kurtosis $\kappa_0$ is a very poor predictor of the coarse-grained kurtosis $\kappa_{final}$ (Fig. \ref{figkappakappa}).

\section{Statistical procedures}
All the results are expressed as mean $\pm$ standard deviation. For statistical comparisons between distributions we used Mann-Whitney U test. We use the convention for representing $p$-values: $p \geq 0.05$ (n.s.), $p<0.05$ (*), $p<0.01$ (**), $p<0.001$ (***), where ``n.s.'' stands for not significant. 
The \textit{p}-values for the comparisons stated in the text are the following:
\begin{itemize}
    \item $\alpha_1$ vs $\alpha_2$:~$\textit{p}$-value$=8\times 10^{-5}$
    \item $\alpha_2$ vs $\alpha_3$:~$\textit{p}$-value$=4\times 10^{-19}$,
    \item $\alpha_1$ vs $\alpha_3$:~$\textit{p}$-value$=8\times 10^{-21}$,
    \item $\beta_1$ vs $\beta_2$:~$\textit{p}$-value$=0.17$,
    \item $\beta_2$ vs $\beta_3$:~$\textit{p}$-value$=3\times 10^{-10}$,
    \item $\beta_1$ vs $\beta_3$:~$\textit{p}$-value$=3\times 10^{-13}$,
    \item $z_1$ vs $z_2$:~$\textit{p}$-value$=6\times 10^{-5}$,
    \item $z_2$ vs $z_3$:~$\textit{p}$-value$=6\times 10^{-6}$, 
    \item $z_1$ vs $z_3$:~$\textit{p}$-value$=3\times 10^{-10}$,
    \item $\mu_1$ vs $\mu_2$:~$\textit{p}$-value$=4\times 10^{-13}$,
    \item $\mu_2$ vs $\mu_3$:~$\textit{p}$-value$=3\times 10^{-8}$,
    \item $\mu_1$ vs $\mu_2$:~$\textit{p}$-value$=3\times 10^{-8}$.
\end{itemize}

\section{Relationship between mean firing rate and scaling}
It is natural to wonder whether the average firing rate could serve as a simple readout of the system’s state.
 Indeed, firing rate varies almost monotonically with $\kappa_{th}$ (Fig.~S5A) within a given experimental condition (active, passive, non natural images phases). 
 This is similar to previous findings for urethane-anesthetized rats, where stronger signatures of criticality are often associated with more synchronized, lower-rate regimes~\cite{castro_and_2024}.

\begin{figure*}[!h]
\centering
\includegraphics[width=\textwidth]{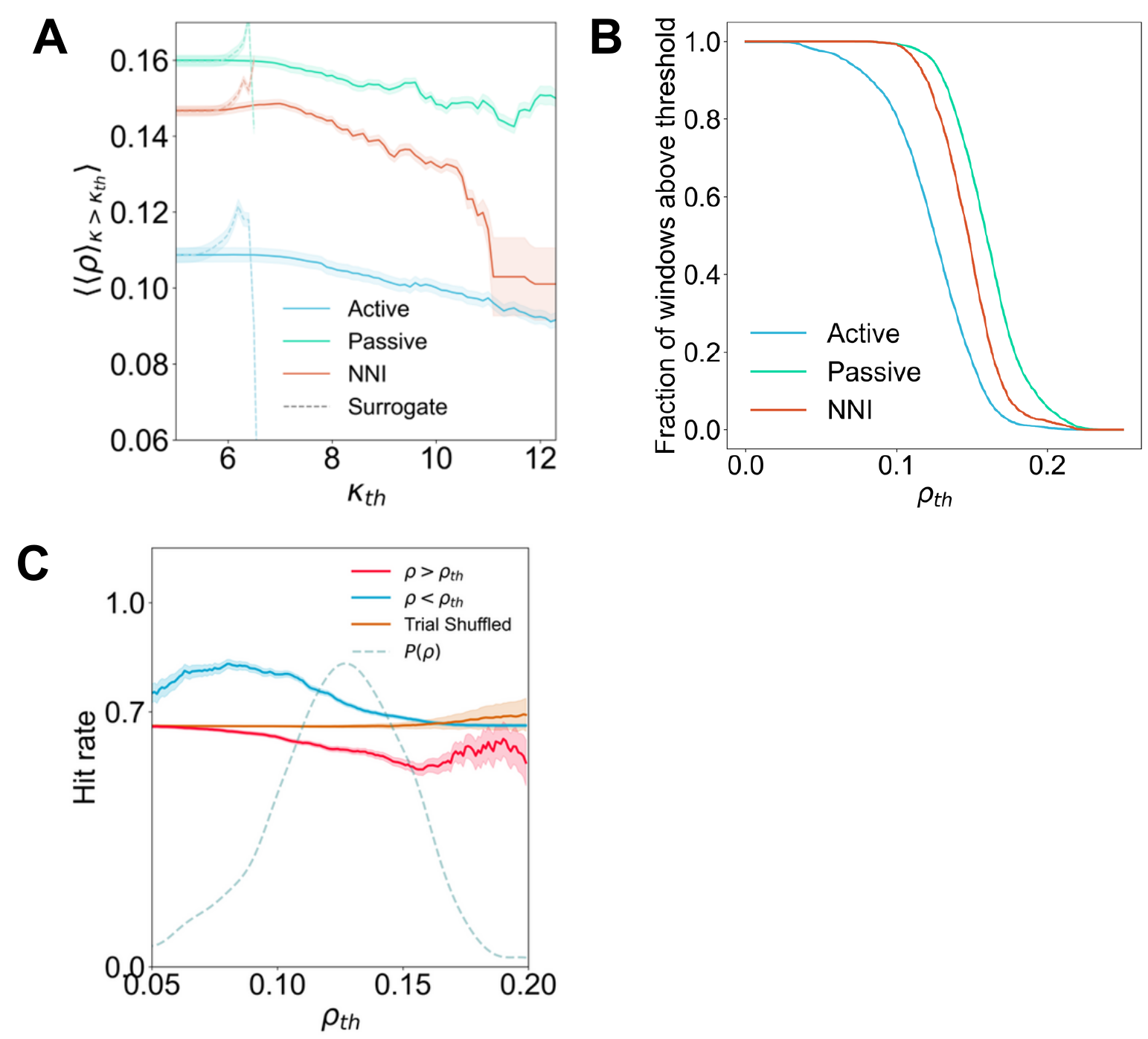}
\caption{\textbf{Firing rate vs threshold.} \textbf{(A)} The firing rate is computed as the average over the windows above a given kurtosis threshold for each mouse, and then averaged across mice. As higher kurtosis is associated with increased synchronous activity, the firing rate is expected to decrease with increasing kurtosis. \textbf{(B)} Complementary cumulative distribution for the firing rate of the three phases. \textbf{(C)} Hit rate is calculated using a firing rate threshold, with results displayed in blue for windows below the threshold and in red for those above. In addition, the trial-shuffled hit rate is derived from windows with a firing rate above the threshold in a shuffled version of the experiment's hits.}
\label{figSMrhoth}
\end{figure*}

However, for non-anesthetized mice, firing rate is not nearly as informative as kurtosis about the underlying dynamical state of the network. 
Figure \ref{figSMrhoth} shows two instances of this issue. 
First, we cannot reproduce the message of Fig.~4A with the firing rate. 
The order of the curves of each phase of the experiment is different in Fig.~4A than in Fig.~\ref{figSMrhoth}A-B. 
On the one hand, on Fig.~4A we have $F_{NNI} < F_{passive} < F_{active}$, for any $\kappa_{th}$, pointing to one of our important findings, namely that signatures of criticality in V1 are modulated by the level of task involvement. 
The mean firing rate, on the other hand, gives no clue about that, since $\rho_{active}~<~\rho_{NNI}~<~\rho_{passive}$, for any $\kappa_{th}$ (Fig.~\ref{figSMrhoth}A).
Correspondingly, the complementary cumulative distribution function of $\rho_{th}$ has the same issue (Fig.~\ref{figSMrhoth}B). 
In the second instance, the hit rate below some threshold $\rho_{th}$ is not a monotonic function of $\rho_{th}$ (Fig.~\ref{figSMrhoth}C), rendering the results harder to interpret.

\section{Average shape of the probability distributions}

\begin{figure*}[!h]
\centering
\includegraphics[width=0.9\textwidth]{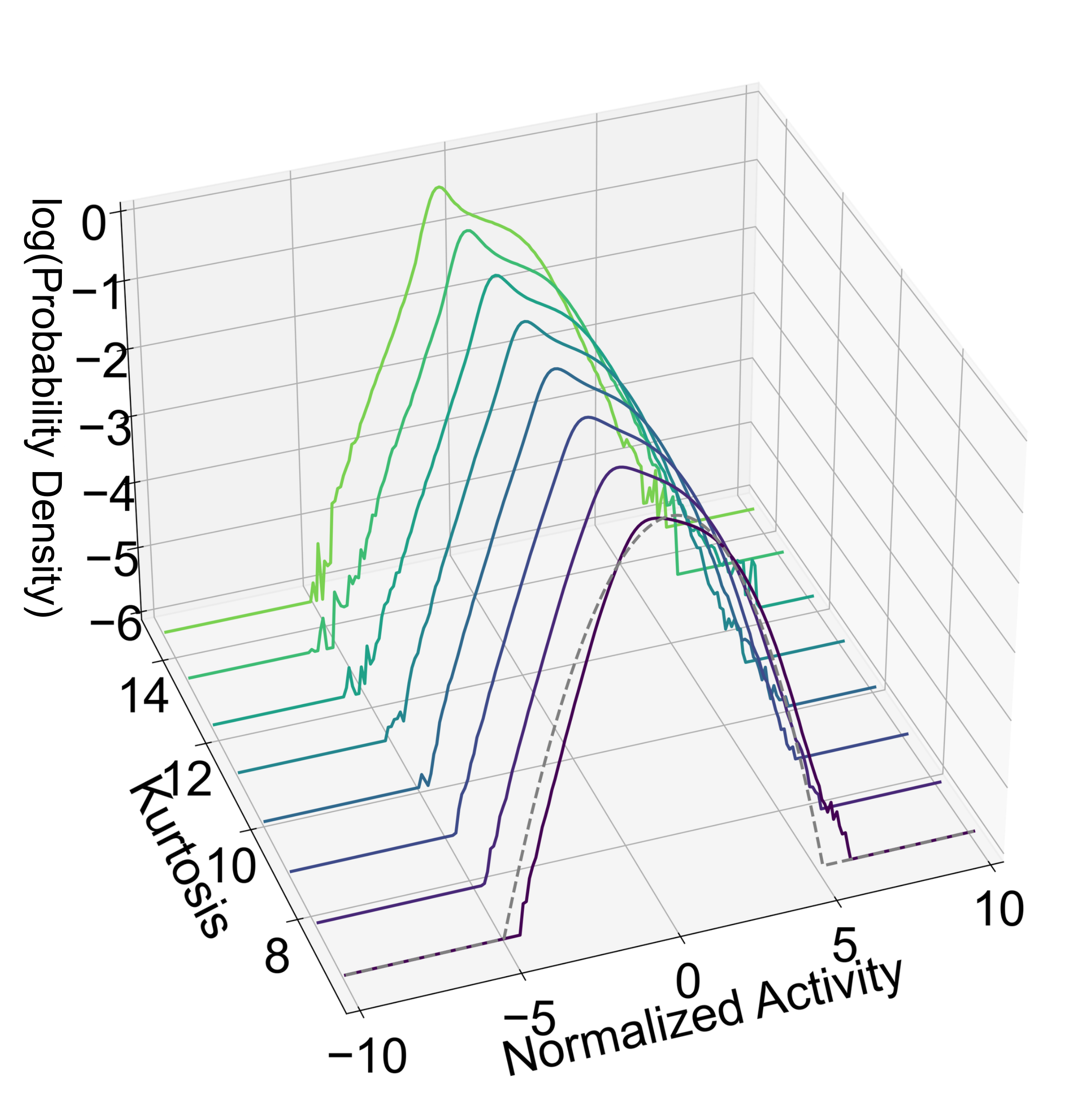}
\caption{\textbf{Mean probability density across different levels of kurtosis.} Although spanning a wide range of kurtosis values, the distributions maintain similar overall shapes. The dashed line represents a Gaussian distribution for reference.}
\label{figSMmultdistcurves}
\end{figure*}

In Figure 2B, we show examples of the probability distribution for the normalized activity of two different windows with markedly different kurtosis values. Since most distributions exhibit non-trivial kurtosis, visually distinguishing them can be challenging. In Figure~\ref{figSMmultdistcurves}, we present the average shape of the normalized activity probability distribution across different kurtosis levels. We find that the distribution shape transitions smoothly away from a Gaussian, developing a pronounced peak slightly below zero. While the extreme cases are easily discernible, the intermediate values -- where the majority of time windows fall (Fig.~3A) -- exhibit only subtle differences. 

\bigskip

\section{Controlling for fatigue effects}

\begin{figure*}[!h]
\centering
\includegraphics[width=0.9\textwidth]{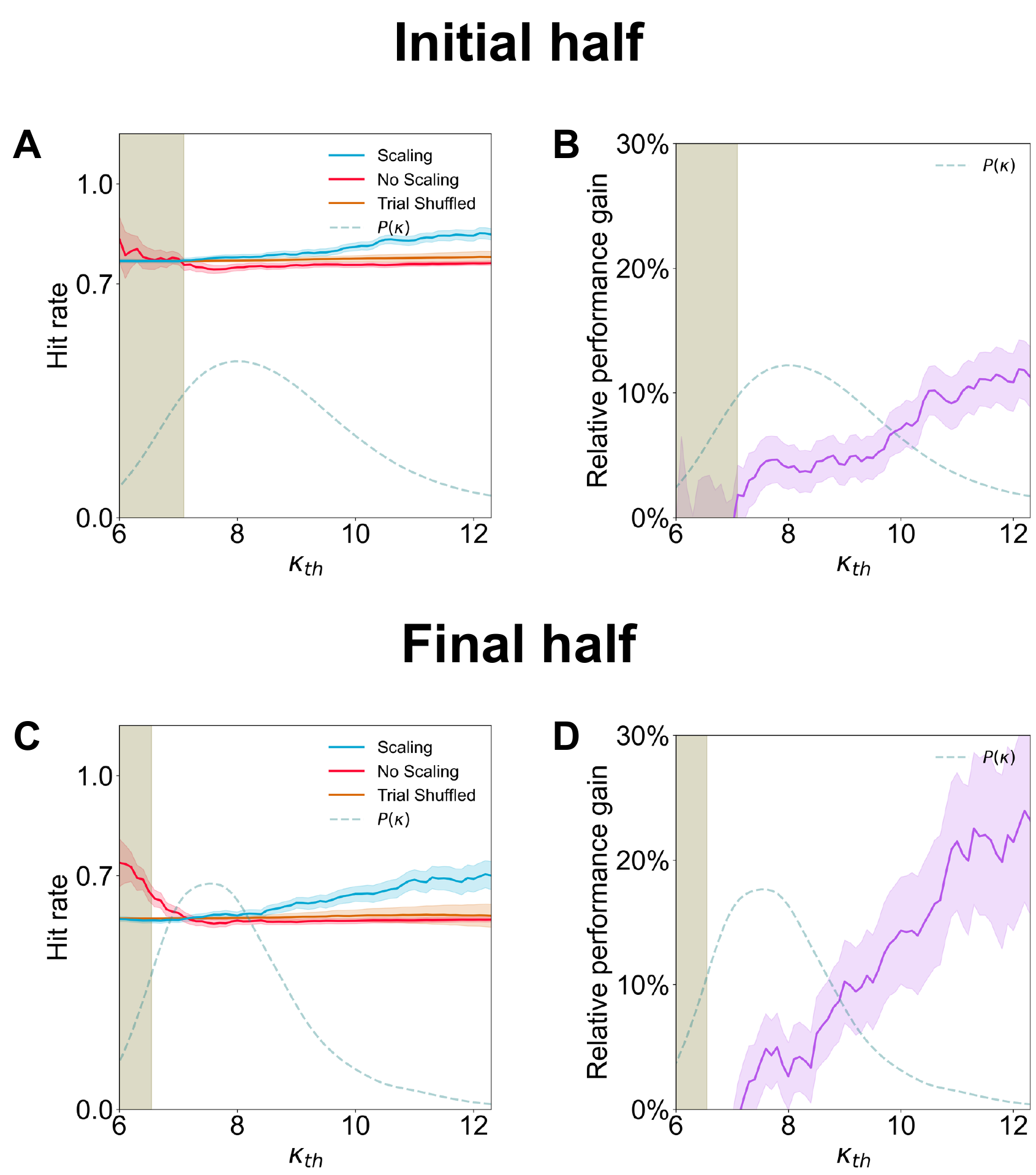}
\caption{\textbf{Hit rate of different parts of the active phase.} \textbf{(A–B)} Hit rate and the corresponding relative performance gain over the first 30 minutes, a period during which the overall hit rate is higher than the one computed over the entire duration of the phase. \textbf{(C–D)} Hit rate and relative performance gain for the last 30 minutes, when overall performance tends to decrease. In both cases, the data reveal a consistent trend: higher kurtosis is associated with increased hit rate, corroborating the observations in the main text.}
\label{figSMfatigue1}
\end{figure*}

To control for the influence of fatigue, we have reproduced our results separately for the first and second half of the active phase (Fig~\ref{figSMfatigue1}). 
What we see is that even if there is a decrease in average hit rate (note the trial shuffled curve in Figs.~\ref{figSMfatigue1}A and~\ref{figSMfatigue1}C), the relationship between hit rate and kurtosis remains.

\begin{figure*}[!h]
\centering
\includegraphics[width=0.9\textwidth]{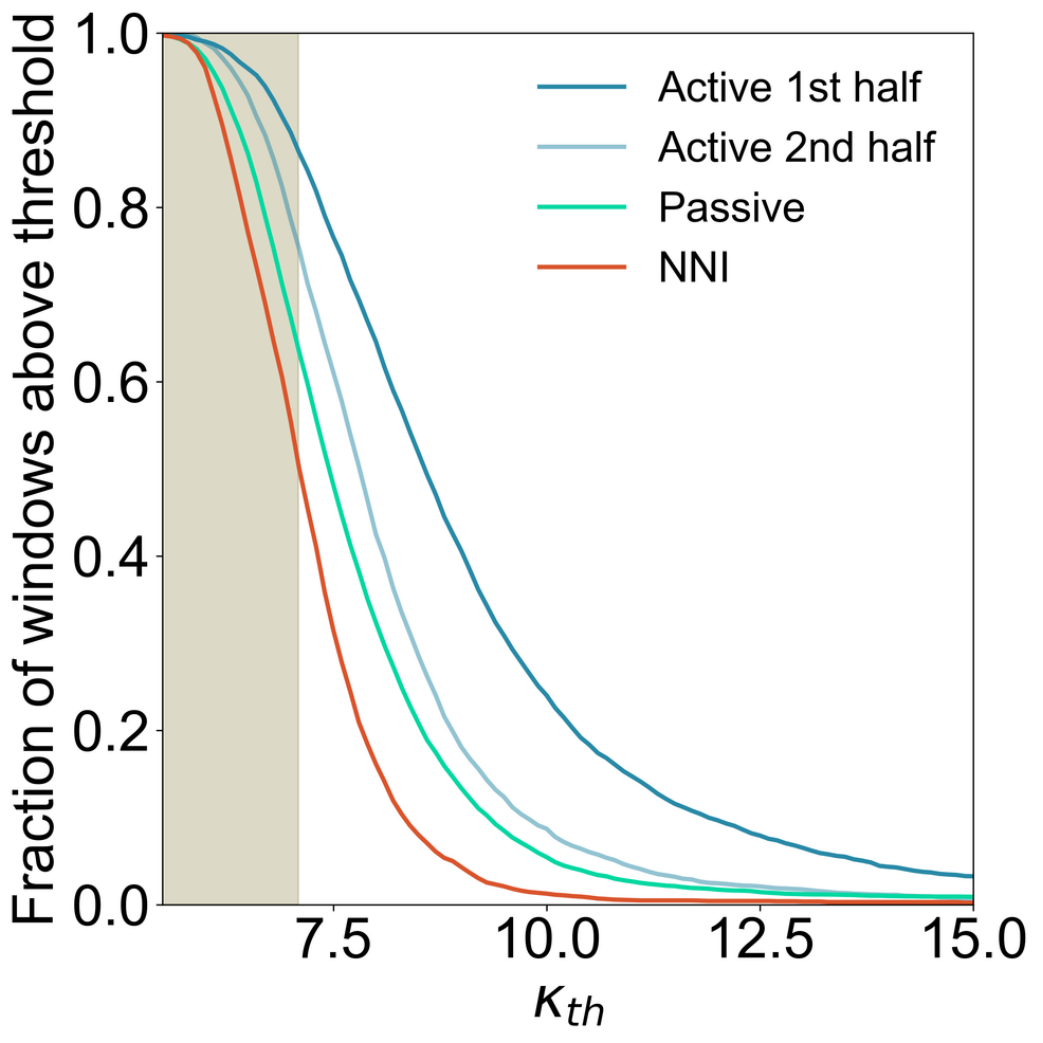}
\caption{\textbf{Fraction of time windows exceeding the kurtosis threshold across all experimental phases.} The active phase – even subdivided into two halves – consistently exhibits higher kurtosis levels compared to the other phases, confirming that the elevated critical signatures persist throughout the active period.}
\label{figSMfatigue2}
\end{figure*}

Additionally, the phase of the experiment where non-natural images are presented shows the lowest kurtosis values, even though it precedes the passive phase (Fig~\ref{figSMfatigue2}). If fatigue were the primary driver, we would expect kurtosis to be lower in later phases due to cumulative exhaustion, yet this is not what we observe.

\bibliography{copelli}